\documentclass{article}

   \usepackage[nonatbib,preprint]{neurips_2024}

\usepackage[utf8]{inputenc} 
\usepackage[T1]{fontenc}    
\usepackage{hyperref}       
\usepackage{wrapfig}
\usepackage{url}            
\usepackage{booktabs}       
\usepackage{amsfonts}       
\usepackage{nicefrac}       
\usepackage[numbers,sort&compress]{natbib}
\usepackage{microtype}      
\usepackage{xcolor}  
\usepackage{microtype}
\usepackage{graphicx}
\usepackage{framed}
\usepackage{caption}
\usepackage{multirow}
\usepackage{pifont}
\usepackage{subfigure}
\usepackage{algorithm}
\usepackage{algorithmic}
\usepackage{fontawesome}
\usepackage{booktabs} 

\usepackage{hyperref}

\usepackage{amsmath}
\usepackage{amssymb}
\usepackage{asymptote}
\usepackage{enumitem}
\usepackage{pifont}
\usepackage{mathtools}
\usepackage{amsthm}
\usepackage{xcolor} 
\usepackage{colortbl} 
\usepackage[capitalize,noabbrev]{cleveref}

\theoremstyle{plain}

\theoremstyle{definition}

\theoremstyle{remark}

\usepackage[textsize=tiny]{todonotes}

\hypersetup{
    colorlinks=true, 
    linkcolor=red,   
    citecolor=blue   
}

\title{\raisebox{-0.3em}{\includegraphics[width=0.7cm]{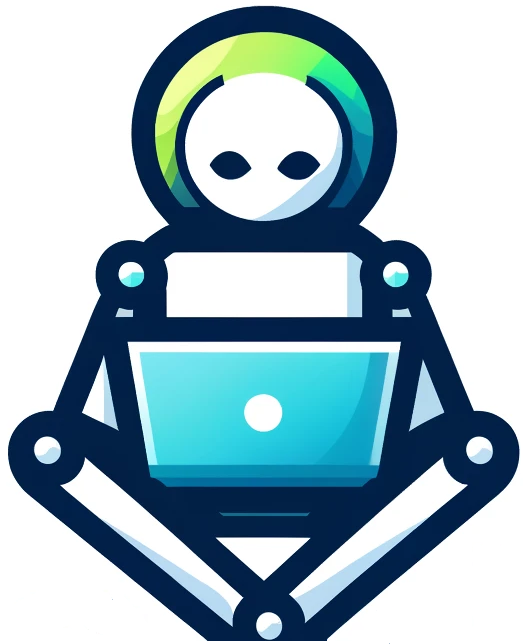}} AutoCoder: Enhancing Code Large Language Model with \textsc{AIEV-Instruct}}

\author{%
Bin Lei$^{1}$ \quad Yuchen Li$^2$ \quad Qiuwu Chen$^2$ \\
$^1$University of Connecticut \quad $^2$AIGCode \\
\texttt{\{bin.lei\}@uconn.edu} \quad \texttt{\{liyuchen, chenqiuwu\}@aigcode.net}\\
}

\begin{document}

\maketitle

\begin{abstract}
We introduce AutoCoder, the first Large Language Model to surpass GPT-4 Turbo (April 2024) and GPT-4o in pass@1 on the Human Eval benchmark test ($\mathbf{90.9\%}$ vs. $\mathbf{90.2\%}$). In addition, AutoCoder offers a more versatile code interpreter compared to GPT-4 Turbo and GPT-4o. It's code interpreter can install external packages instead of limiting to built-in packages. AutoCoder's training data is a multi-turn dialogue dataset created by a system combining agent interaction and external code execution verification, a method we term \textbf{\textsc{AIEV-Instruct}} (Instruction Tuning with Agent-Interaction and Execution-Verified). Compared to previous large-scale code dataset generation methods, \textsc{AIEV-Instruct} reduces dependence on proprietary large models and provides execution-validated code dataset. The code and the demo video is available in \url{https://github.com/bin123apple/AutoCoder}.
\end{abstract}

\section{Introduction}
Code generation is a critical aspect of modern software development. It significantly enhances development efficiency and quality by increasing productivity, reducing errors, standardizing code, accelerating prototyping, and supporting complex systems~\cite{li2024deveval, li2023large,buscemi2023comparative}.Recently, Large Language Models (LLMs), such as GPT-4~\cite{chatgpt2024} and CodeQwen1.5~\cite{codeqwen2024}, have achieved significant advancements on code generation. These models have shown high accuracy in producing code that meets user requirements and have been widely adopted in real-world software development. 

Training large language models requires extensive high-quality data~\cite{hoffmann2022training}. This is particularly crucial for code generation tasks that demand high accuracy~\cite{chen2021evaluating}. OpenAI once hired people to help annotate the Code Instruct dataset for training their InstructGPT~\cite{ouyang2022training}. However, manually annotating large-scale code instruction datasets is both economically and time-consuming~\cite{xu2022ide}. To address this challenge, previous work has employed various automated code annotation methods, such as \textsc{Self-Instruct}~\cite{wang2022self}, \textsc{Evol-Instruct}~\cite{luo2023wizardcoder}, and \textsc{OSS-Instruct}~\cite{wei2023magicoder}. \textsc{Self-Instruct} enhances LLMs' instruction-following capabilities by using strong teacher models to generate synthetic coding instructions for fine-tuning weaker student models. \textsc{Evol-Instruct} improves LLMs' coding abilities by iteratively increasing the complexity of seed code instructions through various heuristics. \textsc{OSS-Instruct} generates diverse and realistic coding problems by drawing inspiration from open-source code snippets. The essence of these methods lies in distilling the knowledge of a powerful teacher model (such as GPT-4 Turbo) to guide a smaller model. This leads to a problem: \textbf{While the small model can achieve significant performance improvements, the final accuracy of the small model is unlikely to surpass that of the teacher model.} Because both the correct and incorrect knowledge from the teacher model are transferred to the small model. Moreover, although using closed-source models reduces costs compared to manual annotation, the cost of using closed-source models remains high. According to our tests, even with the relatively cheaper GPT-4 Turbo model, generating an average of 6,500 high-quality entries for the code instruction dataset costs \$1,000.

\textbf{This raises two questions:}
\begin{enumerate}
[label=\textbf{\arabic*.}]
    \item \textit{\textbf{Can we correct the incorrect knowledge generated by the teacher model to provide more accurate code for the student model?}}
    \item \textit{\textbf{Instead of relying on expensive closed-source teacher models, can we enable our student model to learn autonomously?}}
\end{enumerate}

To address the \textbf{first issue}, we designed a new large-scale code instruction dataset annotation method called \textsc{AIEV-Instruct}. It is an interaction system comprising two agents: a \textit{\textbf{questioner}} and a \textit{\textbf{programmer}}. These agents interact to simulate the process of \textit{\textbf{programmers}} constructing code according to project requirements and conducting unit tests. In each dialogue round, we extract the code generated by the \textit{\textbf{programmers}} and execute it. The execution results are returned to the \textit{\textbf{questioner}} to inform the next round of questions. This process continues until the \textit{\textbf{programmers}}'s code passes the unit tests, ensuring the accuracy of the generated code dataset.

 To address the \textbf{second issue}, we sperate \textsc{AIEV-Instruct} into two stages: the \textit{Teaching Stage} and the \textit{Self-learning Stage}. In the \textit{Teaching Stage}, we rely on proprietary large models as agents for code annotation, similar to previous methods. Once our model surpasses the proprietary models in accuracy on the test set, we transition to the \textit{Self-learning Stage}. In this stage, we use our own model as the agent for code annotation. For detailed methodology, refer to Section~\ref{sec:AIEV-Instruct}.
\begin{figure}[h]
    \centering
    \begin{minipage}[t]{0.49\textwidth}
        \centering
        \raisebox{-\height}{\includegraphics[width=\textwidth,height=5cm,keepaspectratio]{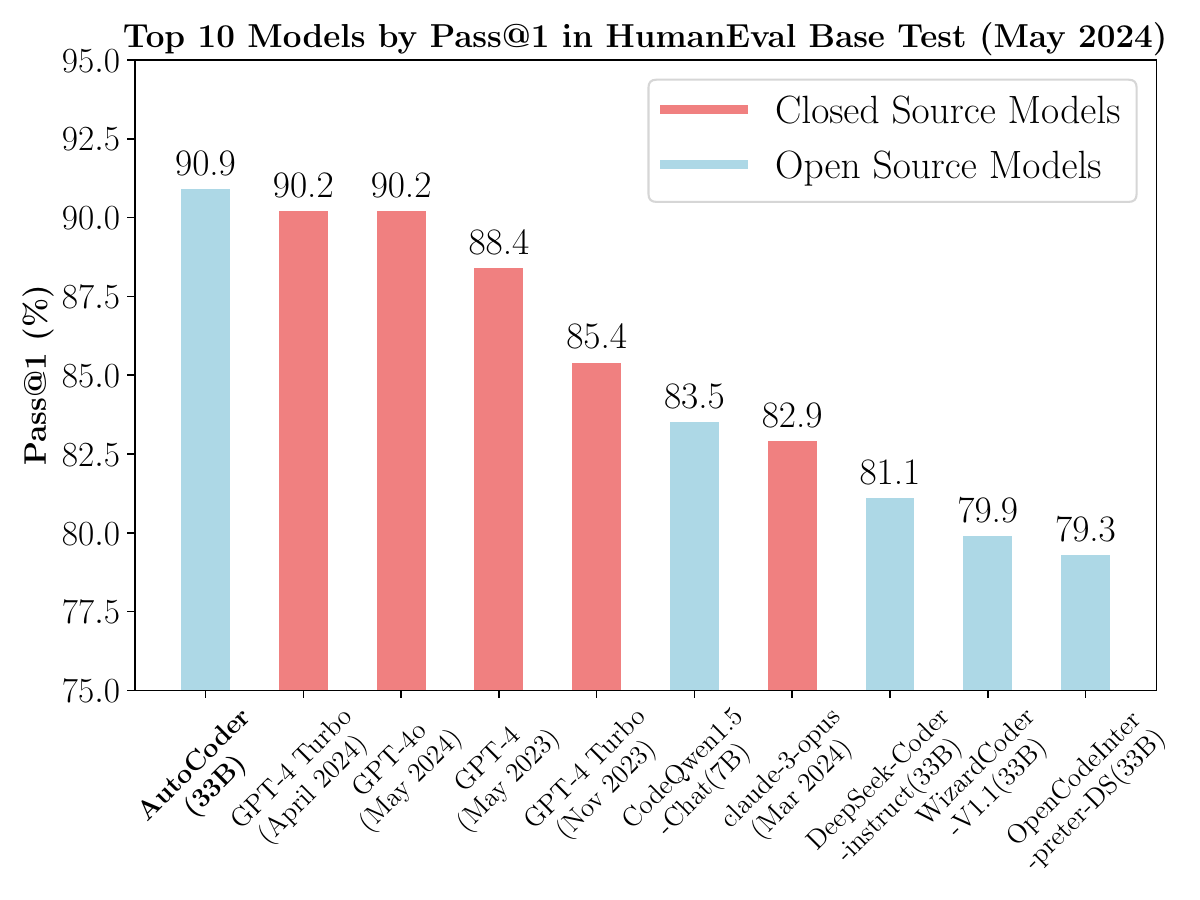}}
        \caption{Pass@1 ($\%$) comparison of Various LLMs on the HumanEval Base Test.}
        \label{fig:Overall_performance}
    \end{minipage}\hfill
    \begin{minipage}[t]{0.49\textwidth}
        \centering
        \raisebox{-\height}{\includegraphics[width=\textwidth,height=5cm,keepaspectratio]{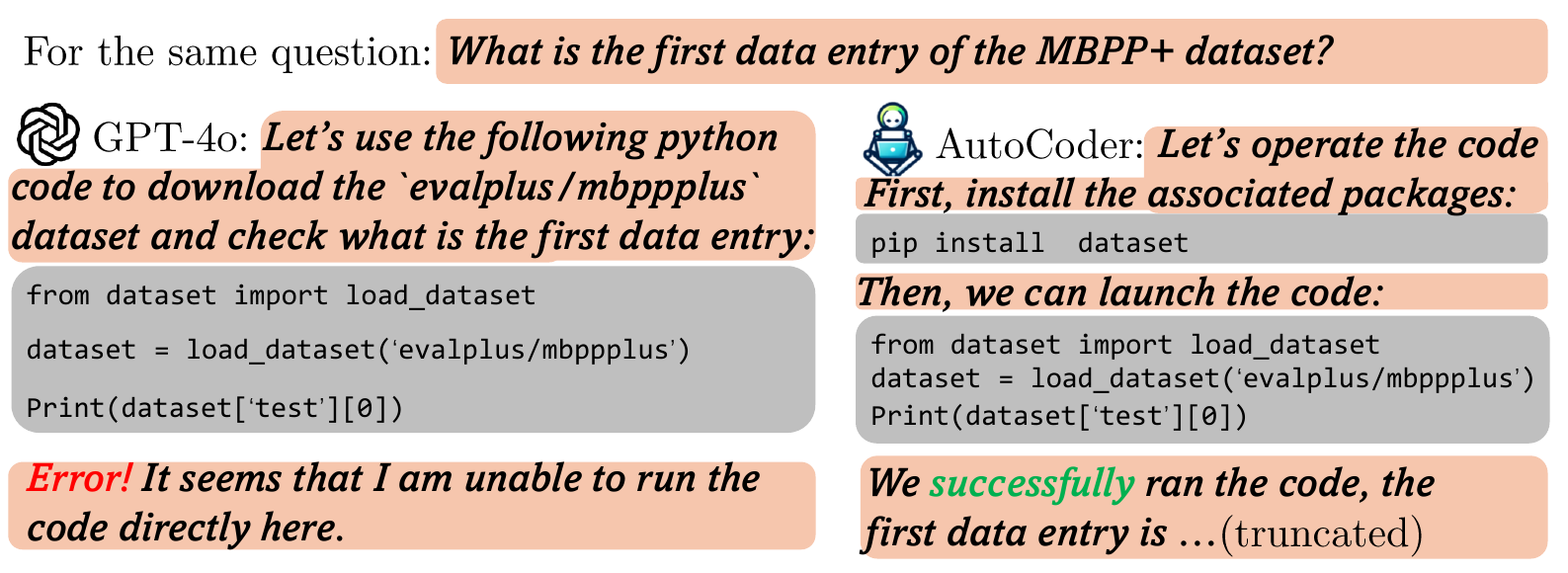}}
        \caption{Comparison of Code Interpreter Functions between AutoCoder and GPT-4o. \includegraphics[height=1em]{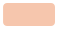}: Nature language generated by the model;\includegraphics[height=1em]{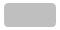}: Code generated by the model. AutoCoder  can recognize \textbf{external package} installation commands, whereas GPT-4o can only run code that includes built-in packages. The demo video is in \url{https://github.com/bin123apple/AutoCoder} .}
        \label{fig:Code_Interpreter}
    \end{minipage}
\end{figure}

Under the support of \textsc{AIEV-Instruct}, we obtained 169K high-quality code instruction data samples. Using this dataset, we trained the AutoCoder series models, including AutoCoder (33B) and AutoCoder-S (6.7B). As shown in Figure~\ref{fig:Overall_performance}, AutoCoder demonstrates higher accuracy. In the HumanEval Base Test, we compared it with the top ten models on the current (May 2024) EvalPlus Leaderboard~\cite{evalplus2024}. The results indicate that AutoCoder's Pass@1 even surpasses that of the current top-ranked models, GPT-4 Turbo (April 2024) and GPT-4o. 

Moreover, as illustrated in Figure~\ref{fig:Code_Interpreter}, AutoCoder boasts a more versatile Code Interpreter function compared to GPT-4o and GPT-4 Turbo. The Code Interpreter is an external program execution environment that large models utilize to execute the code they deem necessary. While GPT-4o and GPT-4 Turbo can identify the code that needs to be executed, they fail to provide the Code Interpreter with the necessary instructions to install external packages required by the programs. This limitation significantly restricts the capabilities of the Code Interpreter. In contrast, AutoCoder can correctly supply the Code Interpreter with the appropriate external package installation instructions, thereby enabling it to execute a wide variety of code.

To comprehensively evaluate the capabilities of AutoCoder, we tested it on several datasets: HumanEval~\cite{chen2021evaluating}, HumanEval+~\cite{liu2024your}, MBPP~\cite{austin2021program}, MBPP+~\cite{liu2024your}, MultiPL-E~\cite{cassano2022multipl}, and DS-1000~\cite{lai2023ds}. To measure the performance improvement of AutoCoder, we compared it to its base model, Deepseek-Coder~\cite{guo2024deepseek}. The results demonstrate that AutoCoder exhibits outstanding performance. As of May 2024, AutoCoder ranks \textbf{1st} among all LLMs on the HumanEval Base Test, \textbf{5th} on the HumanEval Plus Test, and  \textbf{4th} on both the MBPP Base Test and the MBPP Plus Test. Detailed experimental procedures can be found in Section~\ref{sec:Experiment}.

Overall, our contributions are summarized as follows:

\noindent\textbf{We propose \textsc{AIEV-Instruct}}, a novel method for creating high-quality large code datasets. It simulates programmers writing code and conducting unit tests through agent interactions, ensuring annotation accuracy with an external code executor. It includes a \textit{Teaching Stage} and a \textit{Self-Learning Stage}, reducing reliance on expensive closed-source models during the annotation process.
    
\noindent\textbf{We introduce AutoCoder,} a code LLM trained using \textsc{AIEV-Instruct} that excels in code-related tasks. It outperforms top models like GPT-4 Turbo and GPT-4o on the HumanEval benchmark.
    
\noindent\textbf{We enhances the functionality of the current code interpreters.} AutoCoder can provide the code interpreter with the necessary instructions to install external packages, extending the applicability of the code interpreter beyond built-in packages. 
\section{Related Work}
\noindent\textbf{Large Language Models for Code.} Recently, LLMs have shown remarkable abilities in understanding and generating code~\cite{kazemitabaar2023novices}. Trained on extensive datasets covering various programming languages and tasks, these models excel in code completion, bug fixing, and code synthesis~\cite{jin2023inferfix}. Closed-source models like OpenAI's GPT-4~\cite{chatgpt2024}, Claude.ai's Claude~\cite{claudeai2024}, and Google's Gemini~\cite{gemini2024} series have demonstrated superior performance on code tasks. Meanwhile, open-source models specialized for code, such as DeepSeek-Coder~\cite{deepseekcoder2024}, CodeQwen~\cite{codeqwen2024}, Magicoder~\cite{wei2023magicoder}, OpenCodeInterpreter~\cite{zheng2024opencodeinterpreter}, and WizardCoder~\cite{luo2023wizardcoder}, are also emerging. Generally, closed-source models outperform open-source ones due to their larger parameter sizes and broader knowledge base.

\noindent\textbf{Code LLMs Instruction Tuning.} After pre-training large models, we use instruction tuning to optimize them~\cite{gao2020making}, enhancing their ability to understand and execute specific instructions~\cite{chang2024survey}. A major challenge in Instruction Tuning for Code LLMs is the lack of high-quality instruction datasets for code~\cite{rao2024navigating}. Code tasks, such as Text-Code and Code-Code translation, are difficult and time-consuming to annotate manually. OpenAI used human annotators to label various tasks and train InstructGPT~\cite{ouyang2022training}, but they noted that annotating code tasks is prohibitively expensive for large-scale datasets. Since the advent of GPT-4, an increasing number of researchers have leveraged GPT-4 for code annotation to create high-quality instruction tuning datasets. Currently, there are three primary methods: \textsc{Self-Instruct}\cite{wang2022self}, \textsc{Evol-Instruct}\cite{luo2023wizardcoder}, and \textsc{OSS-Instruct}~\cite{wei2023magicoder}.
\textsc{Self-Instruct} boosts LLMs' instruction-following skills by using strong teacher models to generate synthetic coding instructions for fine-tuning weaker student models. \textsc{Evol-Instruct} iteratively enhances LLMs' coding abilities by increasing the complexity of seed code instructions. \textsc{OSS-Instruct} creates diverse coding problems inspired by open-source code snippets. These methods distill the expertise of powerful teacher models like GPT-4 to guide and improve smaller models.
\section{\textsc{AIEV-Instruct}}
\label{sec:AIEV-Instruct}

\subsection{Overall Architecture}
\begin{figure}[ht!]  
\centering 
\includegraphics[width=0.99\textwidth]{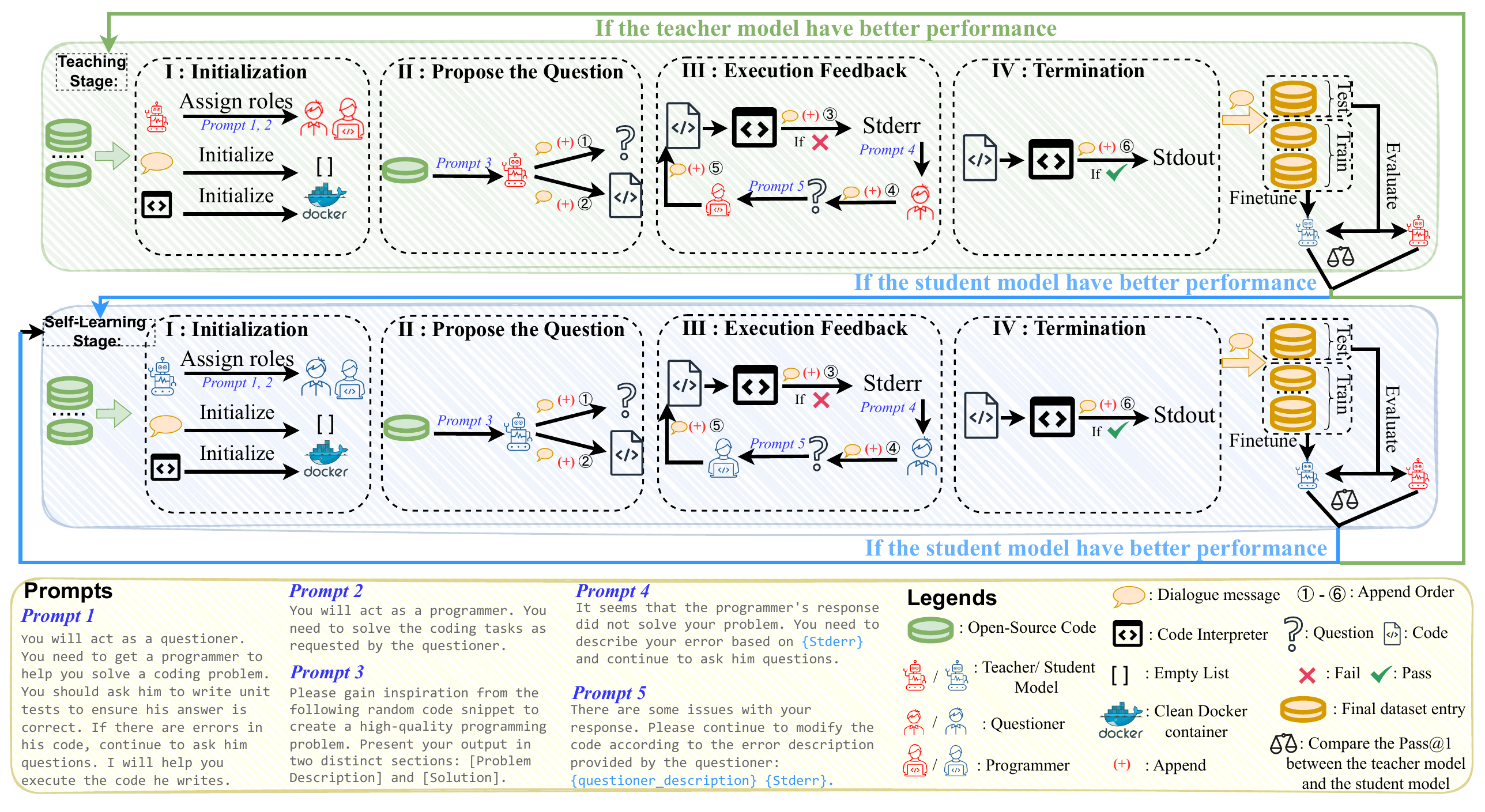} 
\caption{The overall architecture of the \textsc{AIEV-Instruct}.}  
\label{fig:Overall_architecture}
\vspace{-10pt}
\end{figure}

Figure~\ref{fig:Overall_architecture} illustrates the overall architecture of \textsc{AIEV-Instruct}, divided into two stages: the \textit{Teaching Stage} and the \textit{Self-Learning Stage}. In the \textit{Teaching Stage}, the model learns primarily by distilling knowledge from a teacher model. In the \textit{Self-Learning Stage},  it learns autonomously.

In the \textit{Teaching Stage}, we obtain open-source code snippets and use GPT-4 Turbo as the teacher model to supplement and correct them. The process consists of four main steps. \textbf{In I : Initialization,} we initialize the necessary components. GPT-4 Turbo is assigned two roles:  \textit{\textbf{questioner}} and  \textit{\textbf{programmer}}. It can ensure the generated data is diverse, resulting in a more uniform probability distribution rather than converging to a specific dialogue template. The dialogue messages are initialized as an empty list, which will be used throughout the process to store data. Eventually, this list will contain multiple rounds of dialogue, and the entire conversation will be added as a single data entry to our final dataset. Additionally, we need to initialize a Docker container as our Code Interpreter. This container is responsible for installing the required external packages and executing the code that needs verification throughout the process. \textbf{In II : Propose the question,} we first utilize GPT-4 Turbo to execute OSS-Instruct~\cite{wei2023magicoder}, designing a problem description and a specific solution that includes the code snippet based on the open-source code fragment. The difference here is that we require GPT-4 Turbo to provide some \textbf{Unit Tests}. These \textbf{Unit Tests} further ensure the accuracy of the code in our dataset. The dialogue messages initialized in the previous step are sequentially appended with the problem description (\ding{192}), the solution and the unit tests (\ding{193}).
\textbf{In III: Execution Feedback:}, we use multiple rounds of execution feedback to check the generated code, thereby improving the quality of the dataset. First, we input the code snippet generated in the second step into the Code Interpreter. If an execution error occurs, the dialogue messages append the detailed Stderr output (\ding{194}). Meanwhile, this Stderr information is provided to the \textit{\textbf{questioner}}, who will generate a natural language description based on the Stderr. This natural language description is also appended to the dialogue messages (\ding{195}). Next, both the natural language description and the Stderr are provided as new questions to the \textit{\textbf{programmer}}, who will continue to modify the code. The dialogue messages will append the new code it generates (\ding{196}) and repeat this process. \textbf{In IV: Termination}, we also use the Code Interpreter to run the code generated by the \textit{\textbf{programmer}}. If the program executes successfully, the Stdout is appended to the dialogue messages (\ding{197}). This completes the analysis of one data entry.

After analyzing every 2000 data entries, we split the new data into a test set and a training set in a 1:9 ratio. The training set is used to train the student model (AutoCoder). After training, we use the test set to evaluate both the teacher model and the student model. Upon completion of the evaluation, we compare the Pass@1 of the two models. If the teacher model performs better, we continue executing the \textit{Teaching Stage}. If the student model performs better, we move to the \textit{Self-Learning Stage}. The difference between the \textit{Self-Learning Stage} and the \textit{Teaching Stage} is that in the \textit{Self-Learning Stage}, we replace the original teacher model with the student model. The student model itself is assigned as the \textit{\textbf{questioner}} and \textit{\textbf{programmer}} , and it completes the entire execution feedback process.

\subsection{Dataset Analysis}

\noindent\textbf{Dataset Generation.} To prevent \textbf{data contamination} in test sets from resulting in overly high performance on certain benchmark datasets (such as HumanEval), we used code from two datasets that had already undergone contamination detection: \textit{Magicoder-Evol-Instruct} and \textit{Magicoder-OSS-Instruct}~\cite{wei2023magicoder}. We collected a total of 186K original code entries from these two datasets. After de-duplication, we input these data into our AIEV-Instruct pipeline to generate the dataset. We set the maximum number of execution feedback iterations in AIEV-Instruct to 7. If the generated code fails to execute successfully and pass all unit tests after 7 attempts, that data point is discarded. The \texttt{gpt-4-turbo-2024-04-09} is used as the teacher model.

\begin{figure}[ht!]  
\centering 
\includegraphics[width=0.99\textwidth]{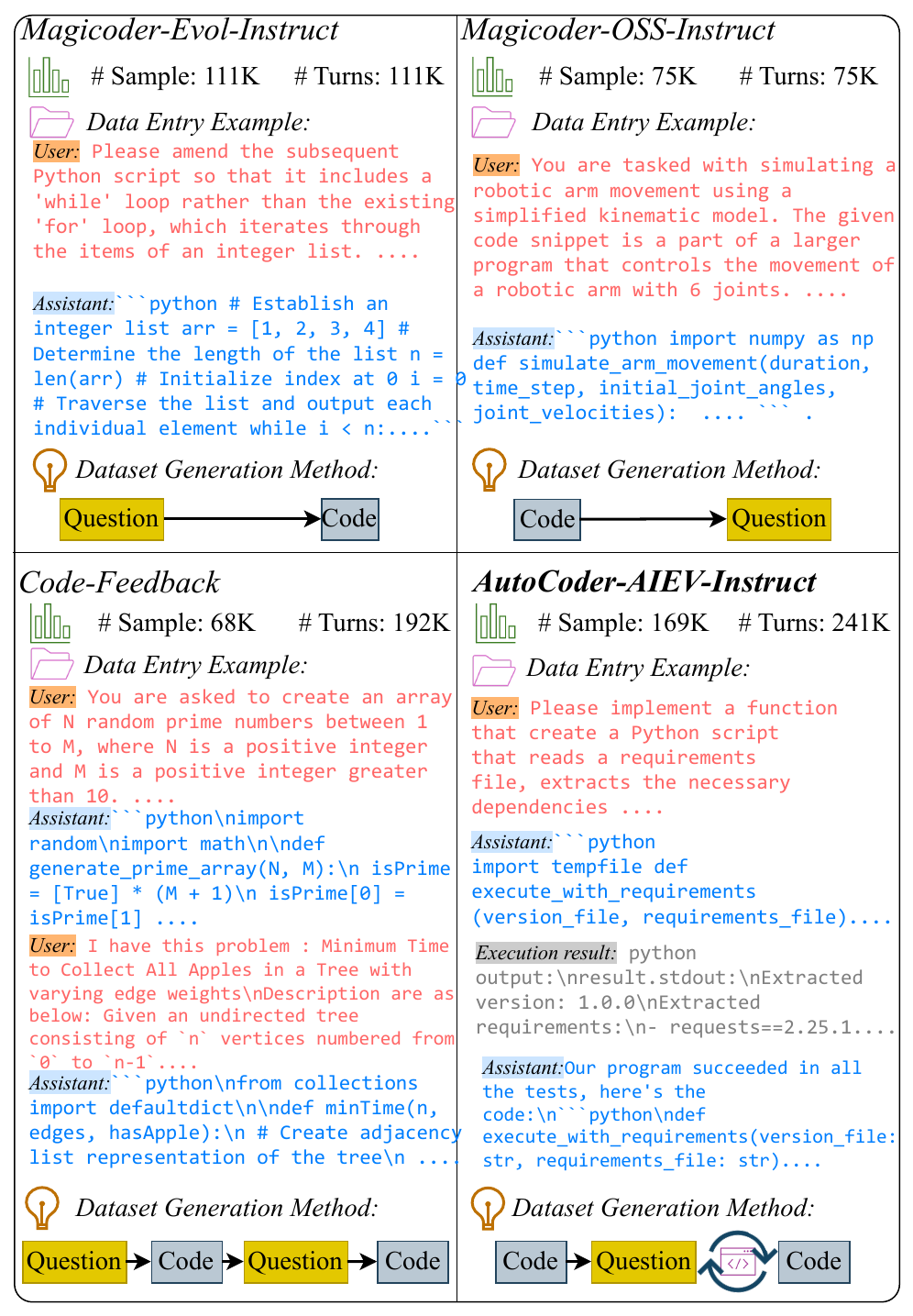} 
\caption{The comparison between the AutoCoder-AIEV-Instruct and other large  code datasets.}  \label{fig:dataset_sample}
\vspace{-15pt}
\end{figure}

\noindent\textbf{Dataset Comparision.} We compared our dataset \textit{AutoCoder-AIEV-Instruct} with several current large code instruction datasets. The comparison results are shown in Figure~\ref{fig:dataset_sample}.
The dataset \textit{AutoCoder-AIEV-Instruct} contains 169K data samples, totaling 241K rounds of dialogue. Among these, 150K rounds are contributed by multi-round dialogue data samples. Besides including the main function, it also encompasses subsequent package installations, code execution errors, or results, as well as various error analyses. Compared to the original \textit{Magicoder-Evol-Instruct} and \textit{Magicoder-OSS-Instruct}, it adds unit tests, which further enhances the accuracy of code-related tasks. Additionally, compared to \textit{Code-Feedback}~\cite{zheng2024opencodeinterpreter}, it includes more execution feedback results, reducing the multi-round dialogues for code block concatenation and enhancing the coherence of the context.

\noindent\textbf{Dataset Decontamination.} Similar to the data processing method used by StarCoder~\cite{li2023starcoder}, we also performed decontamination for \textit{AutoCoder-AIEV-Instruct}. Specifically, we tested each code snippet from HumanEval, MBPP, DS-1000, and MultiPL-E against every code snippet in \textit{AutoCoder-AIEV-Instruct} using Levenshtein distance. If the similarity exceeded 90\%, the data entry was removed. Through this process, we excluded a total of 113 data entries.

\noindent\textbf{Dataset Accuracy Theoretical 
 Analysis.} \textsc{Evol-Instruct} generates code from questions using a teacher model. Thus, the theoretical maximum accuracy of the \textsc{Evol-Instruct} dataset should closely match the teacher model's accuracy in generating correct code \textit{c} for given problems \textit{p}. Thus, we get $\mathcal{A}_{Evol} \approx \mathcal{P}(\textit{c} \mid \textit{p})$. 
\textsc{OSS-Instruct} generates problems from open-source code using a teacher model. Therefore, the theoretical maximum accuracy of the \textsc{OSS-Instruct} dataset should closely match the teacher model's accuracy in analyzing and interpreting open-source code \textit{c}. Thus, we get $\mathcal{A}_{OSS} \approx \mathcal{P}(\textit{p} \mid \textit{c})$. 
For \textsc{AIEV-Instruct}, we align problem descriptions and code by asking LLMs to add unit tests to the original code. After adding unit tests, we can obtain new code $\textit{c}^*$ and new problem descriptions $\textit{p}^*$. Here, we can make an assumption: $\mathcal{P}(\textit{p} \mid \textit{c}) < \mathcal{P}(\textit{p}^* \mid \textit{c}^*)$. This is because LLMs usually find it easier to align problem descriptions with unit tests. For example, suppose we need to write a program \texttt{is\_prime} to detect prime numbers in a list. It is easy to validate the code by providing a few unit tests such as \texttt{assert is\_prime([2,3,4]) == [2,3]} but ensuring that the corresponding program is entirely correct is not as straightforward. with iterative validation and correction, the probability of correctness improves with each iteration. If the probability of error in each iteration is $1 - \mathcal{P}(\textit{p}^* \mid \textit{c}^*)$, and assuming each iteration is independent, the probability of correctness after $n$ iterations is: $\mathcal{A}_{AIEV} \approx 1 - (1 - \mathcal{P}(\textit{p}^* \mid \textit{c}^*))^n > 1 - (1 - \mathcal{P}(\textit{p} \mid \textit{c}))^n$. 

While prior probability of problems $\mathcal{P}(\textit{p})$ is greater than the prior probability of code $\mathcal{P}(\textit{c})$ and the iteration times $n$ is greater than $1$. We can get: 
\[
\mathcal{A}_{Evol} \approx \mathcal{P}(\textit{c} \mid \textit{p}) = \frac{\mathcal{P}(\textit{p} \mid \textit{c})  \cdot \mathcal{P}(\textit{c})}{\mathcal{P}(\textit{p})}
< \mathcal{P}(\textit{p} \mid \textit{c}) \approx \mathcal{A}_{OSS} < 1 - (1 - \mathcal{P}(\textit{p} \mid \textit{c}))^n < \mathcal{A}_{AIEV}\]
Thus, under these assumptions, the accuracy of \textit{AutoCoder-AIEV-Instruct} should be higher than that of \textit{Magicoder-OSS-Instruct} and \textit{Magicoder-Evol-Instruct}.
\section{AutoCoder} 
\subsection{Code Interpreter}
 Code Interpreter assists the model in debugging and executing code, which is essential for fully automating complex coding, scientific computations, and related tasks. Building a code interpreter requires the model to accurately identify the code blocks it needs to run. Currently, only a few models, like GPT-4 Turbo and InternLM-Chat~\cite{cai2024internlm2}, support code interpreters.
 \begin{wrapfigure}{r}{0.7\textwidth} 
  \centering
  \includegraphics[width=0.68\textwidth]{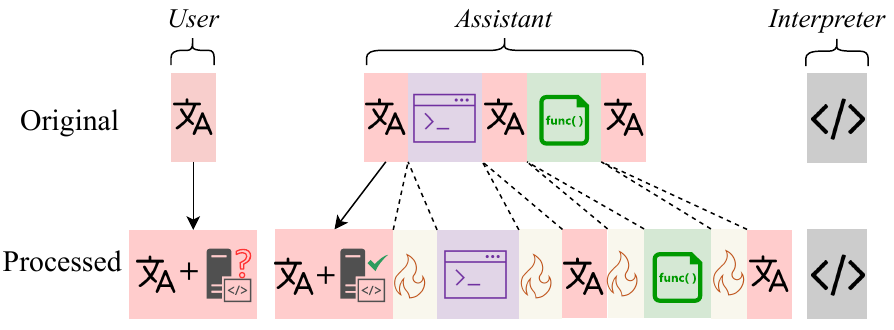}
  \caption{ \textit{AutoCoder-AIEV-Instruct} dataset post-processing.\includegraphics[height=1em]{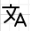}:Nature language;\includegraphics[height=1em]{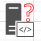}:Code execution request from the User;\includegraphics[height=1em]{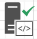}:Code execution request response from the Assistant; \includegraphics[height=1em]{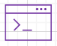}:Bash command;\includegraphics[height=1em]{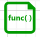}:Code block;\includegraphics[height=1em]{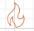}:Special token;\includegraphics[height=1em]{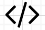}:Execution result.}
  \label{fig:Post_process}
\end{wrapfigure}
 However, a significant limitation of these interpreters is that they operate in a closed environment and cannot interact with external systems, preventing them from executing code that requires external package installations. AutoCoder addresses this issue by enabling the execution of \texttt{bash} commands to install necessary packages. This capability is achieved by teaching the model to run \texttt{bash} commands when appropriate. To facilitate this, we need to perform some post-processing on the \textit{AutoCoder-AIEV-Instruct} dataset.

As shown in Figure~\ref{fig:Post_process}, 
for a simple single execution feedback example, the original data entry contains three parts: natural language from the \textbf{User}; natural language + \texttt{bash} command + natural language + code block + natural language from the \textbf{Assistant}; execution result from the \textbf{code interpreter}.

In the post-processing stage, we mix the natural language of the Code execution request into the User's natural language, enabling the model to correctly learn when to execute the code. Then, we mix the code execution request response into the Assistant's response, so it can generate coherent answers. Finally, we add special tokens before and after the \texttt{bash} commands and code blocks in the Assistant's original response, allowing the model to learn to correctly identify the \texttt{bash} commands and code blocks that need to be executed.

\subsection{Training}

We fine-tuned two base models, Deepseek-Coder 6.7B and 33B, using the \textit{AutoCoder-AIEV-Instruct} dataset to obtain our AutoCoder 33B and AutoCoder-S 6.7B. We utilized the \texttt{AutoTokenizer} package from the \texttt{transformer} library to add four special tokens to these models to enable the Code Interpreter feature for AutoCoder. For hardware, we used 10 nodes with a total of 40 80GB A100 GPUs on a Simple Linux Utility for Resource Management (SLURM) cluster. The NVIDIA Collective Communications Library (NCCL) handled communication between GPUs. In terms of training parameters, we used the ZeRO-Stage 3 feature from the \texttt{deepspeed} library to partition model parameters, with a batch size of 8 per GPU, a gradient accumulation step of 4, a learning rate of 5e-5, and bf16 as the parameter type. The max sequence length was set to 5120 and the total epochs was set to 2. We adopted a full-parameter tuning approach to train the model.

\section{Experiment} 
\label{sec:Experiment}
We tested AutoCoder's abilities in Python text-to-code generation, multilingual code generation, and code generation for data science questions, and compared it with the other models. To ensure a fair comparison with other models and reduce experimental randomness, we \textbf{disabled AutoCoder's external code interpreter during the tests} and used greedy sampling.

\subsection{Python Text to Code Generation}
We evaluated AutoCoder using two of the most commonly used code generation benchmarks: HumanEval~\cite{chen2021evaluating} and MBPP~\cite{austin2021program}. HumanEval is widely used to test various state-of-the-art closed-source models, such as GPT-4o~\cite{openai2024gpt4o}, Claude-3-Ops~\cite{anthropic2024claude3}, Gemini Ultra 1.0~\cite{deepmind2024gemini}, and Llama3 400b~\cite{meta2024llama3}. It contains 164 code generation problems. Compared to HumanEval, MBPP has more test data, with a total of 378 test cases. Additionally, to prevent errors due to the insufficient number of test cases for each code problem in the original benchmarks, HumanEval+ and MBPP+~\cite{liu2024your} have added more test cases to the original datasets. The Pass@1 results are shown in Table~\ref{tab:HumanEval_MBPP}.

\begin{table}[ht]
  \centering
  \caption{Comparison with current SOTA Code Large language models on HumanEval(+) and MBPP(+). The results for GPT-4o, Llama3-400B, and Gemini Ultra 1.0 are sourced from the GPT-4o website~\cite{openai2024gpt4o}. The remaining measurement results are sourced from the Evalplus leaderboard~\cite{evalplus2024}.}
  \label{tab:HumanEval_MBPP}
\begin{tabular}{cccccc}
\hline
                          &                        & \multicolumn{4}{c}{Benchmark (Pass@$1$ $\%$)}                                 \\ \cline{3-6} 
\multirow{-2}{*}{Model}   & \multirow{-2}{*}{Size} & HumanEval     & HumanEval+    & MBPP          & MBPP+         \\ \hline
GPT-4o                    & \faLock                     & $90.2$          & -             & -             & -             \\
GPT-4-Turbo               & \faLock                     & $90.2$          & $\mathbf{86.6}$ & $85.7$          & $\mathbf{73.3}$ \\
Claude 3 Opus             & \faLock                      & $84.9$          & $77.4$          & $\mathbf{89.4}$ & $\mathbf{73.3}$ \\
Llama3                    & $400B$                   & $84.1$          & -             & -             & -             \\
Gemini Ultra 1.0          & \faLock                     & $74.4$          & -             & -             & -             \\ \hline
OpenCodeInterpreter-CL    & $70B$                    & $76.2$          & $70.7$          & $73.0$            & $61.9$          \\
CodeLlama-Instruct        & $70B$                    & $72.0$            & $65.2$          & $75.4$          & $61.7$          \\ \hline
DeepSeek-Coder-instruct   & $33B$                    & $81.1$          & $75.0$            & $80.4$          & $70.1$          \\
WizardCoder-V1.1          & $33B$                    & $79.9$          & $73.2$          & -             & -             \\
OpenCodeInterpreter-DS    & $33B$                    & $79.3$          & $73.8$          & $80.2$          & $68.5$          \\
speechless-codellama-v2.0 & $34B$                    & $77.4$          & $72.0$            & $73.8$          & $61.4$          \\ \hline
OpenCodeInterpreter-CL    & $13B$                    & $77.4$          & $73.8$          & $70.7$          & $59.2$          \\
starchat2-v0.1            & $15B$                    & $73.8$          & $71.3$          & $74.9$          & $64.6$          \\
starcoder2-instruct-v0.1  & $15B$                    & $67.7$          & $60.4$          & $78.0$            & $65.1$          \\
WizardCoder-15B-V1.0      & $15B$                    & $56.7$          & $50.6$          & $64.3$          & $54.2$          \\ \hline
CodeQwen1.5-Chat          & $7B$                     & $83.5$          & $78.7$          & $79.4$          & $69.0$            \\
OpenCodeInterpreter-DS    & $6.7B$                   & $77.4$          & $72.0$            & $76.5$          & $66.4$          \\
Artigenz-Coder-DS         & $6.7B$                   & $75.6$          & $72.6$          & $80.7$          & $69.6$          \\
DeepSeek-Coder-instruct   & $6.7B$                   & $74.4$          & $71.3$          & $74.9$          & $65.6$          \\ \hline
\rowcolor[HTML]{C0C0C0} 
AutoCoder-S               & $6.7B$                   & $78.7$          & $72.0$            &     $79.4$          &       $69.8$        \\
\rowcolor[HTML]{C0C0C0} 
AutoCoder                 & $33B$                    & $\mathbf{90.9}$ & $78.0$            & $82.5$          & $70.6$          \\ \hline
\end{tabular}
\end{table}

Experimental results show that AutoCoder-33B achieved a Pass@1 of $90.9\%$ on the HumanEval benchmark, surpassing all current SOTA code LLMs. On HumanEval+, it achieved a Pass@$1$ of $78\%$, second only to GPT-4 Turbo and CodeQwen1.5-Chat. In the MBPP and MBPP+ tests, its Pass@1 was $82.5\%$ and $70.6\%$ respectively, ranking just below the two large closed-source models GPT-4 Turbo and Claude 3 Opus.
Additionally, despite having only 6.7B parameters, AutoCoder-S still shows impressive performance. It achieved $78.7\%$ and $72\%$ on HumanEval and HumanEval+, respectively. In the MBPP and MBPP+ benchmarks, it achieved $79.4\%$ and $69.8\%$. On MBPP+, its performance is second only to DeepSeek-Coder-instruct (33B) within the 70B parameter level.

\subsection{Multilingual Code Generation}
To test AutoCoder's capabilities in multilingual code generation, we used MultiPL-E benchmark~\cite{cassano2022multipl} to evaluate its performance in six additional commonly used languages. Since MultiPL-E's official library does not support testing closed-source models, we ensured consistent experimental conditions by comparing only with well-known open-source models. The results are shown in Table~\ref{tab:Multi_code}.

The experimental results show that AutoCoder performed exceptionally well in Java, C++, and Rust, achieving $61.4\%$, $68.9\%$, and $60.8\%$ Pass@1 respectively. In the other three languages, its performance was only surpassed by a few models such as CodeQwen1.5-Chat. This demonstrates AutoCoder's robust capabilities in multilingual code generation.

\begin{table}[ht]
  \centering
  \caption{Performance (Pass@1 \%) of AutoCoder on the MultiPL-E benchmark.}
  \label{tab:Multi_code}
\begin{tabular}{cccccccc}
\hline
                        &                        & \multicolumn{6}{c}{Programming Language}        \\ \cline{3-8} 
\multirow{-2}{*}{Model} & \multirow{-2}{*}{Size} & Java & JavaScript & C++  & PHP   & Swift & Rust \\ \hline
Wizard-CL               & $34B$                    & $44.9$ & $55.3$       & $47.2$ & $47.2$  & $44.3$  & $46.2$ \\
CodeLLAMA               & $34B$                    & $40.2$ & $41.7$       & $41.4$ & $40.4$  & $35.3$  & $38.7$ \\
CodeLLAMA-Instruct      & $34B$                    & $41.5$ & $45.9$       & $41.5$ & $37$    & $37.6$  & $39.3$ \\
Deepseek-Coder-Instruct & $33B$                    & $53.8$ & $67.7$       & $63.3$ & $54.7$  & $51.3$  & $54.4$ \\
OpenCodeInterpreter-DS  & $33B$                    & $60.1$ & $69.6$       & $67.1$ & $59.6$  & $54.4$  & $60.2$ \\ \hline
StarCoder-Base          & $15B$                    & $28.5$ & $31.7$       & $30.6$ & $26.8$  & $16.7$  & $24.5$ \\
StarCoder               & $15B$                    & $30.2$ & $30.8$       & $31.6$ & $26.1$  & $22.7$  & $21.8$ \\
WizardCoder-SC          & $15B$                    & $35.8$ & $41.9$       & $39.0$   & $39.3$  & $33.7$  & $27.1$ \\ \hline
CodeLLAMA               & $7B$                     & $29.3$ & $31.7$       & $27.0$   & $25.1$  & $25.6$  & $25.5$ \\
Magicoder-CL            & $7B$                     & $36.4$ & $45.9$       & $36.5$ & $39.5$  & $33.4$  & $30.6$ \\
MagicoderS-CL           & $7B$                     & $42.9$ & $57.5$       & $44.4$ & $47.6$  & $44.1$  & $40.3$ \\
CodeQwen1.5-Chat        & $7B$                     & $58.6$ & $\mathbf{75.7}$       & $65.2$ & $\mathbf{68.9}$  & $\mathbf{58.8}$  & $51.1$ \\ \hline
\rowcolor[HTML]{C0C0C0} 
AutoCoder-S             & $6.7B$                   & $55.7$ & $65.2$       & $62.7$ & $59.6$  & $41.1$  & $50.6$ \\
\rowcolor[HTML]{C0C0C0} 
AutoCoder               & $33B$                    & $\mathbf{61.4}$ & $68.9$       & $\mathbf{68.9}$ & $63.4$ & $53.8$  & $\mathbf{60.8}$ \\ \hline
\end{tabular}
\end{table}

\subsection{Code Generation for Data Science}

We tested AutoCoder's ability to generate code to solve data science problems using the DS-1000 dataset. The DS-1000 dataset contains 1000 questions that require the use of seven commonly used Python data science libraries. We tested all the models using the \textit{completion} mode in DS-1000.

As shown in Table~\ref{tab:DS_1000} the AutoCoder's Pass@1 on Matplotlib-related questions even surpassed that of GPT-4 Turbo. Overall, it is the only model besides GPT-4 to achieve an overall Pass@1 exceeding $45\%$. 
This demonstrates AutoCoder's excellent capability to generate code for data science problems..
\begin{table}[ht]
  \centering
  \caption{Performance (Pass@1 \%) of AutoCoder on the DS-1000 dataset. plt: Matplotlib, np: NumPy
, Pd: Pandas, Py: PyTorch, Scp: Scipy, Sk: Sklearn
, TF: TensorFlow. The result of GPT-4 Turbo \texttt{2024-04-09}, GPT-3.5 Turbo \texttt{0125} and Codex-002 are from the Offical Github of DS-1000~\cite{ds1000}.}
  \label{tab:DS_1000}
\begin{tabular}{cccccccccc}
\hline
                        &                        & $155$        & $220$   & $291$    & $68$      & $106$   & $115$     & $45$         & $1000$    \\ \cline{3-10} 
\multirow{-2}{*}{Model} & \multirow{-2}{*}{Size} & plt & np & Pd & Py & Scp & Sk & TF & Overall \\ \hline
GPT-4 Turbo             & \faLock                      & $72.3$       & $\mathbf{61.8}$  & $\mathbf{42.3}$   & $\mathbf{50}$      & $\mathbf{50}$    & $\mathbf{50.4}$    & $\mathbf{53.3}$       & $\mathbf{53.9}$    \\
GPT-3.5 Turbo           & \faLock                      & $65.8$       & $32.7$  & $30.2$   & $36.8$    & $39.6$  & $40$      & $42.2$       & $39.4$    \\
Codex-002               & \faLock                      & $57$         & $43.1$  & $26.5$   & $41.8$    & $31.8$  & $44.8$    & $39.3$       & $39.2$    \\ \hline
DeepSeek-Coder-Instruct             & $33B$                      & $61.3$       & $50.0$  & $30.9$   & $35.3$      & $36.8$    & $45.2$    & $40.0$       & $42.8$    \\
OpenCodeInterpreter-DS          & $33B$                      & $39.4$       & $57.7$  & $28.2$   & $47.1$    & $40.6$  & $49.6$      & $42.2$       & $42.1$    \\
\hline
CodeGen-Mono            & $16B$                    & $31.7$       & $10.9$  & $3.40$    & $7.00$      & $9.00$     & $10.8$    & $15.2$       & $11.7$    \\
StarCoder               & $15B$                    & $51.7$       & $29.7$  & $11.4$   & $21.4$    & $20.2$  & $29.5$    & $24.5$       & $26.0$     \\
WizardCoder-SC          & $15B$                    & $55.2$       & $33.6$  & $16.7$   & $26.2$    & $24.2$  & $24.9$    & $26.7$       & $29.2$    \\ \hline
CodeLlama-Python        & $7B$                     & $55.3$       & $34.5$  & $16.4$   & $19.9$    & $22.3$  & $17.6$    & $28.5$       & $28.0$      \\
WizardCoder-CL          & $7B$                     & $53.5$       & $34.4$  & $15.2$   & $25.7$    & $21.0$    & $24.5$    & $28.9$       & $28.4$    \\
Magicoder-CL            & $7B$                     & $54.6$       & $34.8$  & $19.0$     & $24.7$    & $25.0$    & $22.6$    & $28.9$       & $29.9$    \\
MagicoderS-CL           & $7B$                     & $55.9$       & $40.6$  & $28.4$   & $40.4$    & $28.8$  & $35.8$    & $37.6$       & $37.5$    \\
InCoder                 & $6.7B$                   & $28.3$       & $4.4$   & $3.1$    & $4.40$     & $2.80$   & $2.80$     & $3.80$        & $7.40$     \\ \hline
\rowcolor[HTML]{C0C0C0} 
AutoCoder-S             & $6.7B$                   & $52.9$       & $38.2$  & $31.6$   & $30.9$    & $31.1$  & $39.1$    & $31.1$       & $37.1$    \\
\rowcolor[HTML]{C0C0C0} 
AutoCoder               & $33B$                    & $\mathbf{72.9}$       & $52.7$  & $36.1$   & $26.5$    & $45.3$  & $46.1$    & $42.2$       & $47.2$    \\ \hline
\end{tabular}
\end{table}
\subsection{Comparison with the Base Model}
\begin{figure}[ht!]  
\centering 
\includegraphics[width=0.99\textwidth]{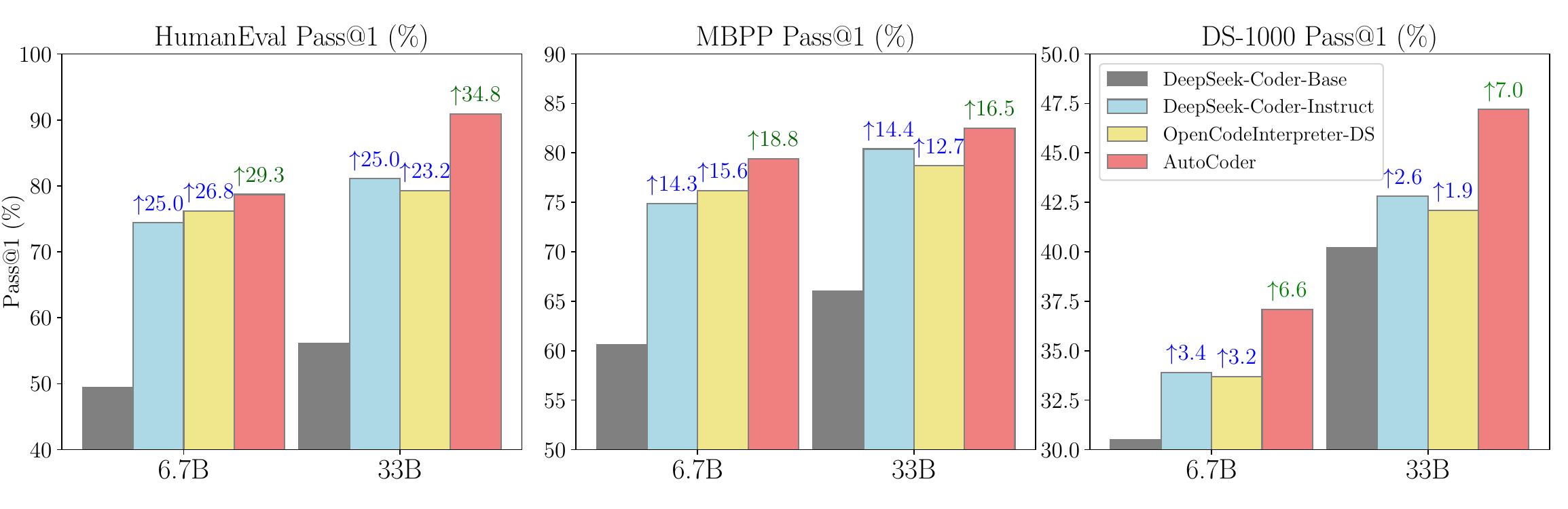} 
\caption{Comparison of AutoCoder with other models sharing the aame base model.}  \label{fig:Compare_with_Base}
\end{figure}

To more intuitively compare the improvements AutoCoder brings to the base model, we compared it with several models trained on the same base model as AutoCoder. As shown in Figure~\ref{fig:Compare_with_Base}, on the HumanEval, MBPP, and DS-1000 datasets, AutoCoder demonstrates a stronger improvement over the base model compared to DeepSeek-Coder-Instruct and OpenCodeInterpreter. Notably, DeepSeek-Coder-Instruct was trained with 2 billion tokens, while AutoCoder achieved better results with only 320 million tokens. This proves the effectiveness of the \textsc{AIEV-Instruct} method.

\section{Conclusion}

We propose \textsc{AIEV-Instruct}, a novel method for creating high-quality code instruction datasets. It simulates programmers writing code and conducting unit tests through agent interactions, ensuring accuracy with execution validation. It includes both a \textit{teaching stage} and a \textit{self-learning stage}, reducing reliance on expensive closed-source models during the annotation process.
Using the dataset generated with \textsc{AIEV-Instruct}, we trained the AutoCoder code LLM. It exhibits excellent performance and surpass the current top models, GPT-4 Turbo and GPT-4o on the HumanEval benchmark. Furthermore, AutoCoder extends the functionality of previous code interpreters by allowing them to automatically install external packages, thus extending the applicability of the code interpreter. Overall, our work provides the community with excellent open-source code large language models and offers new insights for generating high-quality large code instruction dataset.

\bibliographystyle{plainnat}
\bibliography{bio}


\appendix

\end{document}